\documentclass{egpubl}
\usepackage{eg2021}
 
\ShortPresentation      % uncomment for (final) Short Conference Presentation
\usepackage[T1]{fontenc}
\usepackage{dfadobe}  

\usepackage{cite}  % comment out for biblatex with backend=biber
\BibtexOrBiblatex
\electronicVersion
\PrintedOrElectronic
\ifpdf \usepackage[pdftex]{graphicx} \pdfcompresslevel=9
\else \usepackage[dvips]{graphicx} \fi

\usepackage{egweblnk} 
\title{Soft Walks: Real-Time,
      Two-Ways Interaction \\
      between a Character and Loose Grounds}

\author[C. Paliard, E. Alvarado, D. Rohmer, M-P. Cani]
{\parbox{\textwidth}{\centering  
Chloé Paliard$^{1}$, Eduardo Alvarado$^{2}$, Damien Rohmer$^{2}$, Marie-Paule Cani$^{2}$
        }
        \\
{\parbox{\textwidth}{\centering 
        $^1$ LTCI, Télécom Paris, Institut Polytechnique de Paris $\;\,$
       $^2$ LIX, Ecole Polytechnique/CNRS, Institut Polytechnique de Paris
       }
}
}
\begin{document}

 % Testing! Maybe I change the image to show better positions
 \teaser{
 \vspace{-0.5cm}
  \includegraphics[width=1.0\linewidth]{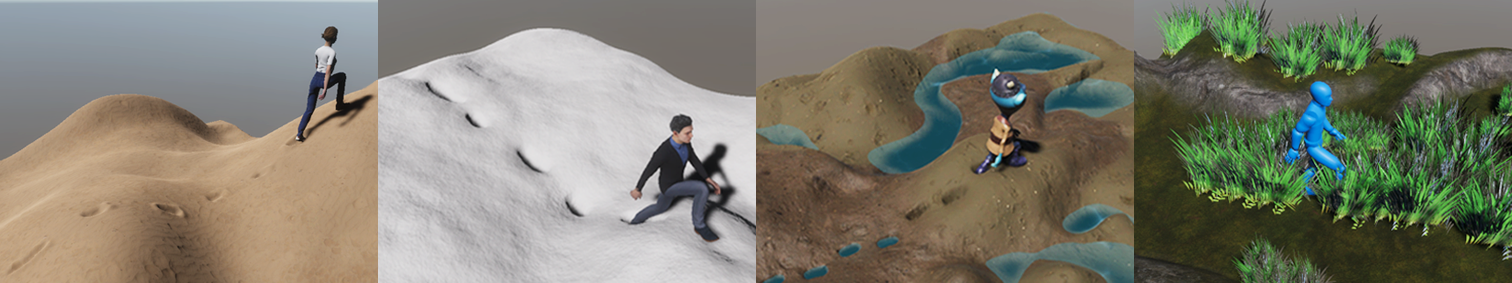}
  \centering
   \caption{
   New layers added to a walking character in a game engine allow to automatically deform a variety of loose terrains and grass, using simple procedural parameters. Changes of gait that are consistent with this evolving environment are generated in real-time.}
   %\vspace{5mm} %30mm vertical space
 \label{fig:teaser}
}

\maketitle

%-------------------------------------------------------------------------

\begin{abstract}

When walking on loose terrains, possibly covered with vegetation, the ground and grass should deform, but the character's gait should also change accordingly. We propose a method for modeling such two-ways interactions in real-time. We first complement a layered character model by a high-level controller, which uses position and angular velocity inputs to improve dynamic oscillations when walking on various slopes. Secondly, at a refined level, the feet are set to locally deform the ground and surrounding vegetation using efficient procedural functions, while the character's response to such deformations is computed through adapted inverse kinematics. While simple to set up, our method is generic enough to 
adapt to any character morphology. Moreover, its ability to generate in real time, consistent gaits on a variety of loose grounds of arbitrary slope, possibly covered with grass, makes it an interesting solution to enhance films and games.

%-------------------------------------------------------------------------
%  ACM CCS 1998
%  (see https://www.acm.org/publications/computing-classification-system/1998)
% \begin{classification} % according to https://www.acm.org/publications/computing-classification-system/1998
% \CCScat{Computer Graphics}{I.3.3}{Picture/Image Generation}{Line and curve generation}
% \end{classification}
%-------------------------------------------------------------------------
%  ACM CCS 2012
% The tool at \url{http://dl.acm.org/ccs.cfm} can be used to generate
% CCS codes.
%Example:

%\ccsdesc[300]{Computing methodologies~Physical simulation}

\printccsdesc
\end{abstract}

%-------------------------------------------------------------------------
\section{Introduction}

Real-time animation of characters walking on loose grounds such as sand, snow, mud or soil, possibly covered with grass, is a key element in video games and virtual-reality applications involving natural environments. Modeling plausible walking motions on such terrains requires two-ways interactions. Indeed, not only should the character's feet deform the loose material of the ground, but this deformation should, in return, dynamically impact the walking style, since both balance and motion IKs are to be adapted to the evolving terrain.

%Such grounds can model terrains that deform under the character's feet such as sand and snow, as well as grounds covers of vegetation. Modeling plausible motion in these grounds requires coupled interaction between the character and its environment. Indeed, the character should adapts its motion to the nature and slope of the prescribed ground he is walking on. However, the terrain itself and the surrounding vegetation should be impacted by the character's presence, which should retro-actively results in a dynamic adaptation of the character's motion. }

While recent work based on deep and/or reinforcement learning allows to animate character motion on terrains with a variety of rigid slopes and obstacles, real-time character animation on loose grounds and deformable vegetation remains a challenge. Indeed, capturing enough learning data would be difficult, while optimizing motion-controllers would require costly training sessions where all collisions and interactions between the character and the environment are fully simulated. These simulations remain notably out of reach when loose material and soft interactions are involved.

%The use of advances in optimization and reinforcement-learning allow recent approaches to model complex dynamic character motion with lots of degrees of freedom in a \dr{fully simulated environment in which interactions are handled using collisions and mechanical constraints [PALvdP18]. Such simulations are well handled in synthetic environments made of rigid obstacles that can be accurately simulated, but become prohibitively costly if deformable material and soft interaction with the character have to be computed.}
% a passer en entat de l'art
%In addition, it provides only indirect control of the resulting motion typically depending on reward functions and training data-set, that are hard to scale easily to video-game like environments, i.e. large variety of terrains and effects that may not be accurately simulated.

This work takes a different path, in proposing a very simple and efficient, yet versatile, method to modify a character's walking style, through two-ways interactions between specific layers of its motion controller and the environment, the latter being consistently deformed through efficient, procedural functions. Since it achieves real-time performances, our method is well suited to game engines such as Unity (Unity Technologies), which we use for illustration.

%The main idea detailed 
Our main contribution is a new, multi-layer animation modifier for the character (Sec.~\ref{sec:layered_model}) 
which applies at different levels of details while reusing and extending standard game-engine principles, namely combination of approximated physics-based dynamics, pre-set motions and inverse kinematics (IK). 
We first introduce, at a high level, a global orientation controller allowing the character to dynamically tilt and improve balance by targeting an equilibrium state that depends on the nature and slope of the terrain. %(Sec.~\ref{sec:controler}).
%, using a single controller targeting an objective equilibrium state .
%control of the character orientation allowing it to tilt and dynamically adapt its balance to the nature and slope of the terrain, using a single controller targeting an objective equilibrium state (Sec.~\ref{sec:controler}). 
Second, at a finer scale, we model two-ways interactions between the character's feet and the environment by using simple distance-based rules used to both dynamically deform the environment and adapt feet placement and character walking style in real-time.
%(Sec.~\ref{sec:feet}).
%to better take into account the terrain characteristics and dynamic deformation.
%The higher level layer is based on 
%a single controller acting on the global orientation of the character, represented as a simple proxy primitive. Thanks to its simplicity this controller can be manually parameterized and dynamically adapted to different types of terrain and slope. The motion of the character limbs can be seen as more refined layers, and we propose in this work the local adaptation of the character's feet position using simple procedural rules to better take into account the terrain characteristics and dynamic deformation.
%demonstrate 
%that enriching these two simple layers along a standard game engine allows to improve the character's walk in 
Lastly, we show how such layered control can be used to enhance
standard game engines in 
different scenarios, such as walking on sand, soil, mud, snow and grass with different slopes (Sec.~\ref{sec:application}). In each case, a specific deformation is applied to the environment, which consistently impacts the walking style.
%character's motion.
%
The method runs in real-time, and opens the way to extensions to other loose environments, as well as refinement in adding more layers to the character's motion controller.

%\dr{Need to reformulate the plan, when it converged}
%In Section 2, we introduce a quick study of the related work in character animation and environment simulation. In Section 3, we explain our idea of layered model and we detail our method in Section 4. In Section 5 and 6 we present some results and possible future work.

%-------------------------------------------------------------------------
\vspace{-0.2cm}
\section{Related Work}

Modeling plausible character motion is a complex problem, which couples behavioral intents with physics-based equilibrium constraints. Fully kinematic models were explored in early work~\cite{SM01} leading to real-time adaptation of walking cycles to uneven terrains, at the price of lacking dynamic response to terrain changes. Coupling preset gaits with dynamic events was 
%typically 
handled through the help of animation controllers~\cite{SIMBICON} 
%associated to character limbs~\cite{SIMBICON}. 
%These controllers 
that convert differences between the actual pose and the prescribed one as torques, which are then fed into a physically-based simulation computing dynamic state changes. 
%evolution of limbs. 
Coupled with reinforcement-learning~\cite{PALvdP18} and/or adapted reward functions~\cite{Kwon2020}, these approaches can model a wide range of plausible complex motions up to acrobatic effects~\cite{Won2020}. These motions automatically adapt to their simulated environment, at the price of their high computational cost and indirect control, and thus cannot be directly applied to loose deforming terrains in standard game engines.
In reverse, gathering and processing large prerecorded motion data-bases allow for the use of deep-learning methods~\cite{HKS17} which, once learned, can achieve real-time performance. Still, the need of an extensive motion data-base would not scale well to the variety of loose environments targeted in this work.

Closer to our goals, simple dynamic models applying forces to the center of mass to control the general balance of a character, seen as an inverted pendulum~\cite{Kwon2010}, allow a good high-level trade-off between efficiency and plausibility. Mitake et al.~\cite{MAA09} proposed a high level representation of an oscillating proxy volume, triggering pre-recorded detailed animations (walking, running, falling). This work achieves real-time animation but is restricted to pre-set animations on flat grounds. Other works attempt to exploit to a greater degree the environmental impact on the dynamic model. Bermudez et al.~\cite{Bermudez2018} proposed a method for applying drag forces to a character walking in a fluid and adapting its motion to the medium. Similarly to ours, this approach rely on high level controller-based model. However, this previous work only take into account the influence of the environment, ie. the fluid, toward the character's gait, but not the coupled two-ways interactions between them as in this work.% In addition, we also introduce a more detailed coupling between the feet position and terrains slope.}

%Closer to our approach, a simple dynamic model has been proposed by Mitake et al.~\cite{MAA09}. Their characters are associated, at a high level, to a proxy solid that moves and oscillates similarly to an inverted pendulum. Once this high level representation is set, a pre-recorded visual model is played based on the type of motion (walking, running, falling). Their approach is fully compatible with real-time application, but is only defined for flat, rigid grounds, and the detail motion is restricted to play existing animations. Our approach is also relying on this high level oscillatory model represented as a torque, but extends it to steep, loose grounds, and introduces a coupling between the feet position and the terrain.

\vspace{-0.2cm}
\section{A layered model for character motion}
\label{sec:layered_model}

Our method takes as input an animated character model from a standard game engine, ie. a mesh geometry, a rigged skeleton with preset animations, an IK system applied to feet bones, on top of a simple proxy-geometry used for collision processing, and a global dynamic behavior computed in simulating a single rigid body model expressed by a center of mass and an inertia tensor.
This model allows the character to walk on flat and sloppy terrains, but does  not allow to dynamically adapt its gait to varying ground slopes, leading to a robotic-looking motion that lacks dynamic movement. Furthermore, it does not model any two-ways interaction with the environment, therefore restricting the animation to rigid grounds.

The aforementioned elements can be seen as a layered model for character motion-control, where each layer acts at a different level of detail. 
They span from global control for the rigid body that controls the position and orientation of the coarse geometric collider, to local control for the IK, which positions, computed using dynamic ray-casting, allow for accurate feet positioning on the ground.
%Indeed, the rigid body acts at the most global level of the general position and orientation, which controls the coarse collider model, acting itself on the feet positions.The latter are then accurately placed on the ground thanks to a ray-cast computation, to finally be used in the IK system to compute the legs bones angles. 
%
In this work, we enhance the two layers we just mentioned, by firstly adding an extra swing-control term to the global rigid body model, and secondly by dynamically adapting feet IKs, enabling them to trigger deformations of the surrounding ground and grass. Taken together, these contributions result in consistent interactions between a walking character and an arbitrary, natural-looking environment.

%handled by the IK system used to finely place the feet position from ray-cast computation, 

\vspace{-0.2cm}
\subsection{A high level controller for tilting and swinging motion}
\label{sec:controler}

\begin{figure*}[htb]
  \centering
  \includegraphics[width=\linewidth]{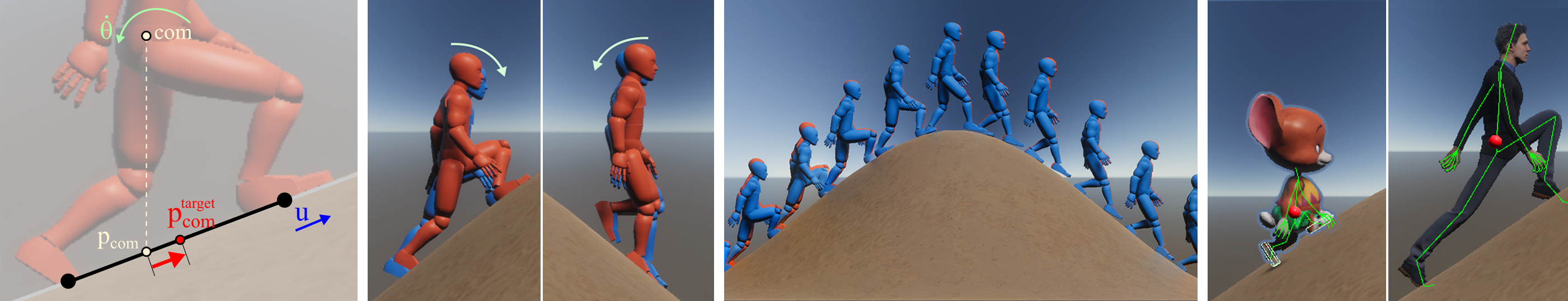}
  \caption{\label{fig:supportPolygon} Left: The torque magnitude depends on the distance between the middle of the support polygon $p_{com}^{target}$ and the projection of the character's center of mass \(p_{com}\) on the terrain, as well as on the current angular velocity \(\dot \theta\). Middle: Comparison between characters without (red) and with (blue) our torque-based controller.  The blue character exhibits an increased swinging motion, which dynamically tilts it %frontwards 
  forward or backward, depending on the slope.  Right: The controller is robust to changes of character morphology, since the oscillating motion automatically adapts  when the center of mass (red sphere) is positioned closer to or further from the ground.}
\end{figure*}

The first layer of our model is a single global controller used to generate a torque applied to the rigid-body. This controller allows the character to self-stabilize on a steep terrain, while adding a dynamic swinging behavior conveying the notion of the effort developed by the character to remain stable and move forward.
While controllers are usually expressed as a difference between the current angles and some objective angles for specific limbs, our high-level controller aims at stabilizing the general character, which does not have a well defined notion of angle in itself. Instead, we propose to use the more relevant notion of center of mass, and in particular its projection along the gravity direction to the support polygon of the character. 
%the projection along the gravity direction of the center of mass to the support polygon of the character.

Since we aim at modeling a tilting motion in the velocity direction, we simplify the representation to a 2D-planar problem, where the plane is defined by the gravity direction, the current character velocity, and is passing through its center of mass. The support polygon is dynamically computed as the segment defined by the projection of the feet positions onto the ground in this plane. The objective projection of the center of mass \(p_{com}^{target}\) is thus expressed as the middle of this segment (see Fig.~\ref{fig:supportPolygon}-left). The torque magnitude is finally computed as a proportional derivative controller:
%expression is computed as a proportional derivative controller such that
%As this support polygon changes when the character climbs a slope, we dynamically compute a target projection of the center of mass \(p_{com}^{target}\) to be placed in the middle of the projection of the feet into the ground along the gravity direction, thus optimizing its balance as being illustrated in Fig.~\ref{fig:supportPolygon}. As we aim at a tilt and balancing motion in the direction of the character's motion, the computation of the torque can be conducted in 1D along this local motion direction. The final torque magnitude expression is computed as a proportional derivative controller such that
%We finally compute our controller as a proportional derivative expression with the torque magnitude expressed as
\begin{equation}
    \tau = \alpha\, (p_{com}^{target}-p_{com})\cdot u + \beta \, \dot{\theta}\mbox{ ,}
    \label{eq:controller}
\end{equation}
where \(p_{com}\) and \(p_{com}^{target}\) are respectively the coordinates of the current and the target projections of the center of mass onto the ground, \(u\) is the unit vector along the terrain slope, and $\dot{\theta}$ is the angular velocity of the rigid body around the axis orthogonal to the 
%represented 
plane.
The control parameter \(\alpha\) is used to 
%constraint 
tune the freedom of the character to tilt out of its ideal equilibrium position, while \(\beta\) sets how much of dynamic swinging is allowed.
The values of these parameters can easily be adapted to the nature of the environment as explained in Sec.~\ref{sec:application}. 
%In our applications, we considered the typical values \(\alpha=6\) and \(\beta=30\) for all characters, 
%\question{we should add something to express the length: we are mixing absolute length difference with angles, we should normalize this length by some basic value - like saying that we expressed all lengths in their respective meter system, and your character is about 1.6 to 1.8m height, or anything that allows people to reproduce or scale these values with their own length system.}
As shown on Fig.~\ref{fig:supportPolygon}-middle and right, this new high-level controller automatically adapts to various slopes and character morphologies,
%while using a single value for 
without requiring any parameter tuning. To ensure the accuracy of foot positioning and limit any sliding artifact, a small offset is added on the character with the PD controller to match the IK-objective of the animation.
%\(\alpha\) nor \(\beta\).

%\begin{figure}[htb]
%  \centering
%  \includegraphics[width=\linewidth]{images/PD_soft_blend.png}
%  \caption{\label{fig:pdComparison}
%           TODO.}
%\end{figure}

%\begin{figure}[htb]
%  \centering
%  \includegraphics[width=\linewidth]{images/PD_steep_blend.png}
%  \caption{\label{fig:pdComparison}
%           TODO.}
%\end{figure}
           
%\begin{figure}[htb]
%  \centering
%  \includegraphics[width=\linewidth]{images/cover_test_1.png}
%  \caption{\label{fig:pdComparison}
%           TODO.}
%\end{figure}

%\begin{figure}[htb]
%  \centering
%  \includegraphics[width=\linewidth]{images/PD_multiple_characters_comp_horizontal.png}
%  \caption{\label{fig:pdComparison}
%           \question{To be clarified + explained (or removed and we only show different character at the end if we don't have anything very meaningfull to say)} \ea{The CoM for each character is at different positions. This makes each of them to have different balancing behaviors, such that taller characters tilt and oscillate more than shorter characters, although the controller works robustly in all of them}.}
%\end{figure}

\subsection{Two-ways interactions between feet and loose grounds}
\label{sec:feet}

The second, new control layer models both the deformation of the environment and its reciprocal action on the character's gait.
%%\dr{a modified?} IK system setting the positions of the feet approaching the ground, together with a mechanism modeling the action-reaction on/from their immediate environment.
%%% MP: Attention: nos deux layers sont deux niveaux
%%% d'un controleur de mouvement (ça ne peut pas être
%%% une position)
%notably acting and reacting with their immediate environment. 
%The relative \mpc{\bf [MP: Why "relative"? The position seems to be computed in a global frame, and not as a relative position in some local frame?]} 
To accurately compute the contact between the foot and the ground, which can be deformed dynamically, our method relies on the use of ray-casting, emitted vertically from the foot to the ground. Once the contact point is computed, it serves as an IK key to the associated leg.
%to the local deformation of the terrain, our method relies on the use of a real-time vertical ray-casting from the foot to the ground
%To support slopes, the IK key controlling ground positioning for the next support foot is computed by casting a vertical ray from a point above the foot.
%This enables feet positions to automatically adapt
%to local changes of ground height, with back-propagation to the leg, impacting walking style.
%and back-propagates to the character gait using IK.
%\cp{Our method enhances this well-known system by adding }%
This approach is complemented by
mechanisms modeling the loose nature of the ground, and possibly some soft volumetric layer (eg. grass) above it.

\paragraph*{Interaction with the ground.}
% surface 
% MP: the ground is a surface

The ground is represented as a height-field grid on which feet dynamically imprint. 
We inspire from Sumner et al.~\cite{Sumner99} for procedurally computing footprints: The geometry of the next support foot 
%in contact with the height field 
hitting the ground carves it locally, while  removed material is added on the side of the footprint.
We use the following three scalar parameters to adapt the deformation to different types \(t\) of grounds: 
\begin{description}
%\textit{Depth coefficient} \(d_t\) - 
\item[The depth coefficient \(d_t\)] defines the height subtracted in one frame to each grid-cell of the terrain in collision with the feet. Thus, smaller values correspond to harder terrains. 
%\textit{Compression coefficient} \(c_t\) -
\item[The compression coefficient \(c_t\)]
describes the amount of compression the ground material can bear, and therefore controls the volume of material to be moved to the edge of the print (eg. mud is compressible, as some water will flow out of the footprint under pressure, while dry sand is not). In practice, the height of the grid-cells surrounding the feet is increased by \(c_t\).
\item[The smoothness coefficient \(s_t\)]
%\textit{Smoothness coefficient} \(s_t\) - 
tunes the aspect of ground deformation. It corresponds to the magnitude of a Gaussian filter applied to the deformed height-field. A high value models material such as dry sand, where fine grains lead to a smooth appearance.
%capacity of material to be compressed without global volume preservation leading visually to more or less material displaced at the edge of the footprint. Material such as dry sand will exhibit incompressible behavior while mud will not, as the water will flow out of the footprint under pressure.
\end{description}

% \mpc{ \textbf{ [MP: Add an equation for height field deformation? Values are given in Sec 4, without the equations to use them!] }}
 
\paragraph*{Interaction with loose surrounding elements.}
\begin{figure}[htb]
  \centering
  \includegraphics[width=0.75\linewidth]{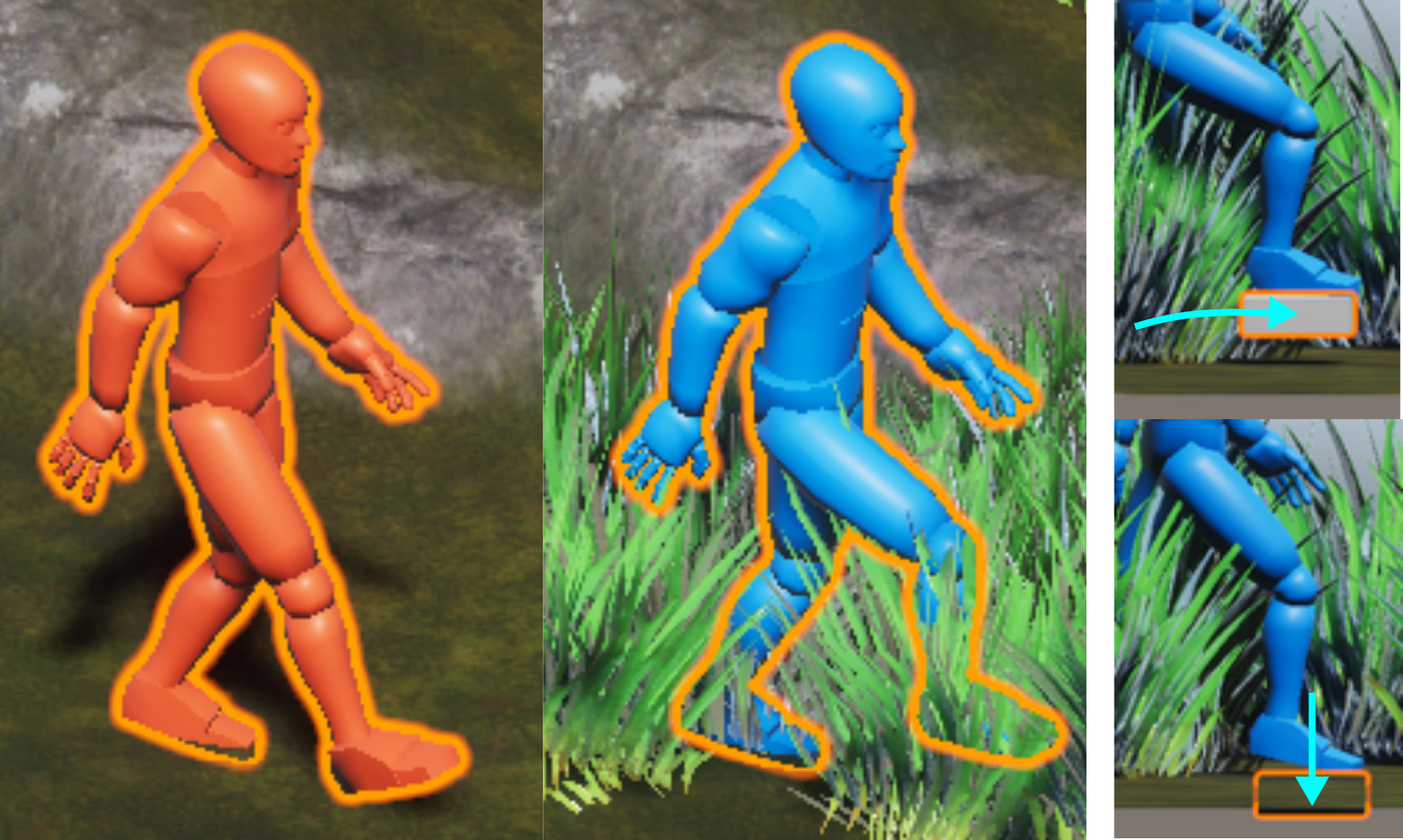}
  \caption{\label{fig:vegetation} 
  We use a moving platform (right) acting as a fake target ground for the foot, but suddenly moving down when hit, to model the presence of grass and its ability to be crushed. \vspace{-0.5cm} }
\end{figure}

%Additionally to the ground represented as a surface, 
In addition to loose grounds, surrounding elements defined in a volumetric layer above it,
%dense volume around the feet 
such as short or dense vegetation, act as loose material and should be impacted, and change the character motion in return.
%represent an important aspect of the environment also reacting as a loose material. 

The action of the feet on all geometric elements in such layer is handled using a procedural deformation,
%environmental elements 
 %function depending on 
function of the distance to the walking foot: 
%We choose to model the general loose behavior of these elements as 
We choose to apply a translation \(t\) with linearly decreasing magnitude around the foot position $p_f$, computed at a point $p$ as:
%such that for any point $p$ in space
\begin{equation}
    t(p) = \mbox{clamp}\left(t_{max}-\gamma \,\|p-p_f\|\right)\;\;\frac{\;\;\;\left(p-p_f\right)_{down}}{\|p-p_f\|}
    \mbox{ ,}
\end{equation}
where \(t_{max}\) and \(\gamma\) are respectively control parameters for the magnitude and influence area of the translation, and \(v_{down}\) 
%corresponds to 
denotes a modified vector that always points downward, computed as: \(v_{down} = (v_x,-|v_y|,v_z)\), where \(y\) is the 
%gravity 
%% MP: non, vertical, c'est opposé à la gravité?
vertical direction.

Secondly, surrounding elements should directly affect 
characters' walking styles. One may observe on footage of natural gaits within 
%in the middle of small 
short, dense vegetation, that humans tend to raise their knees higher and land more abruptly when reaching a short and dense vegetation layer.
%vegetation level. 
Indeed, walkers cannot accurately perceive the actual distance from their feet to the ground,
%real floor, 
but only roughly estimate it. They thus typically target higher positions,
%supposed ground 
while the vegetation will suddenly crush under their feet.
%grass under his feet may still continue 
%to sink. 
We capture this using
%employing 
virtual platforms 
%temporary false floors 
placed under the feet, and modeling the perceived ground (see Fig.~\ref{fig:vegetation}). 
These platforms, serving as targets for the IK system, are procedurally controlled depending on the height of the soft layer (eg. vegetation).
%vegetation. 
%When the character moves a foot up, 
During the upward foot motion phase, the associated platform progressively rises and uplifts the foot. During downward motion,
%the down motion phase, 
it is set to a specific, fixed height. 
Lastly, when the foot reaches it, the platform either falls down quickly or just disappears (depending on the targeted material for the above-ground layer), leading to a sudden adjustment of the character feet onto the actual ground.

\vspace{0.2cm}
\section{Application to various terrains}
\label{sec:application}

%All our results were 
We implemented our method within the Unity Game Engine\footnote{\url{https://unity.com}}, running in real-time on standard laptops, and used it to model 
interactions between a walking character and a variety of terrains,
%approach to model 
%the interaction with different types of terrains, 
namely dry sand, soil, mud and snow. We considered $10 \times 10$ meters terrains with a maximum height of $10$ meters and character sizes ranging from $1$ to $1.9$ meters. Height fields were typically sampled using $512^2$ points. 

\begin{figure}[htb]
  \centering
  \includegraphics[width=0.9\linewidth]{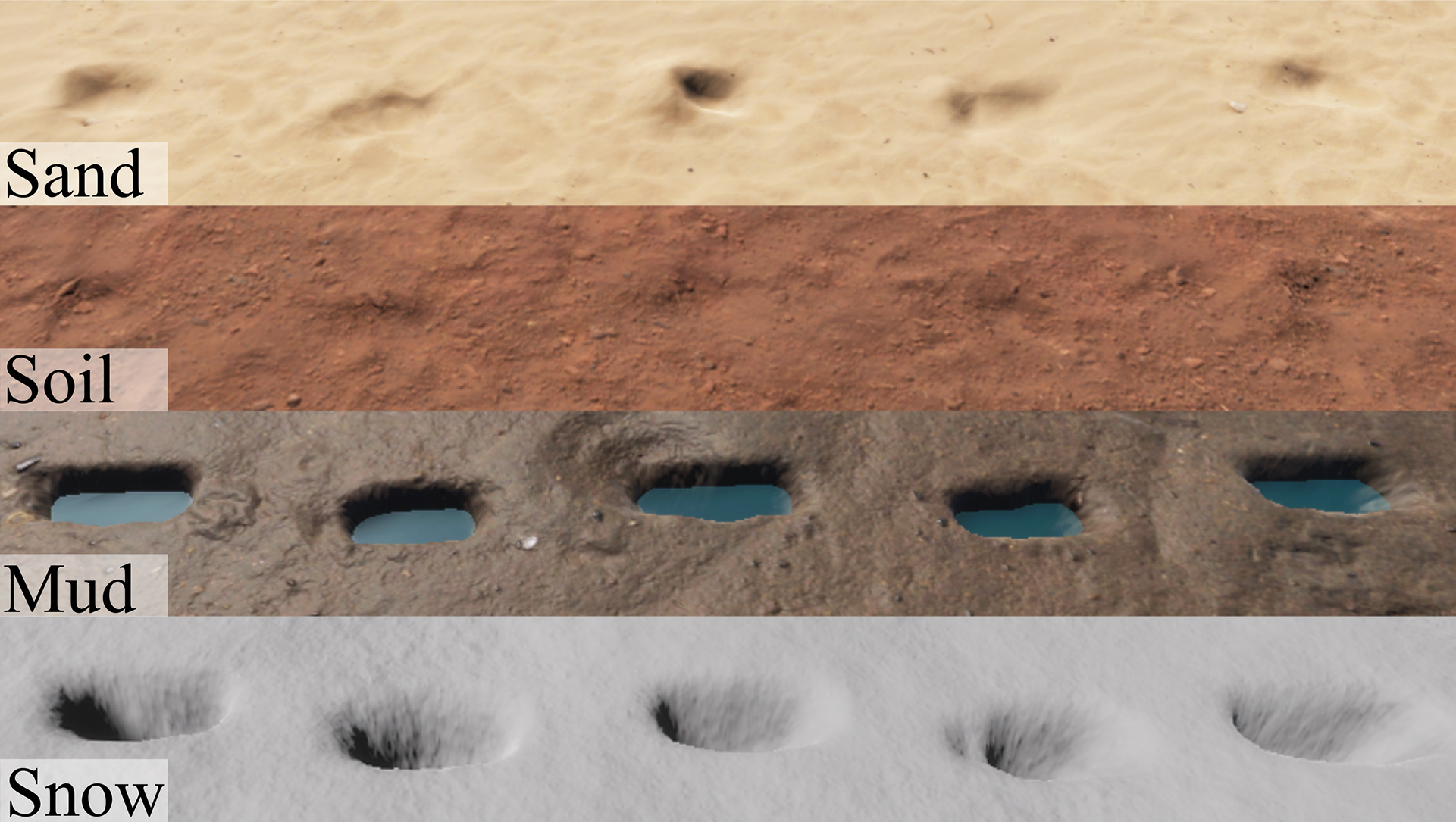}
  \caption{\label{fig:footprintsTerrains}
   Footprints on different terrains. }
\end{figure}

Footprints computed on terrains without vegetation are illustrated in Fig.~\ref{fig:footprintsTerrains}. Controller parameters in Eq.~(\ref{eq:controller}) were set to \(\alpha=30\), and \(\beta=6\). 
%These effects are parameterized 
The variety of prints was achieved by tuning the depth, compression, and smoothness parameters described in Section~\ref{sec:feet}. Dry sand was set to show limited %compression, 
depth changes,
while exhibiting non-compressible behavior and large smoothing. Soil was set to smaller depth but higher compression
%compressibility 
than sand. Mud and snow were set to allow 
%the feet to 
deeper carving, 
%in more depth 
prints on mud being set to be sharper that those on snow.
%. While snow exhibits very smooth footprints, mud is associated to a sharper trace. 
The resulting instantaneous change of terrain height -- controlled through the depth coefficient -- has an immediate effect on the character's gait, as shown on the joint video.
%depends on the geometrical deformation of the terrain associated to the depth coefficient.

\vspace{-0.5cm}
%
%\begin{table}[h]
%\centering
%\begin{tabular}{ |c|c|c|c|c| } 
%\hline
%Terrain & $d_t$ & $s_t$ & $c_t$ \\
%\hline
%Sand & 0.5 & 3 & 0.2 \\
%Soil & 0.4 & 2 & 0.6 \\
%Mud & 2 & 2 & 0.6 \\
%Snow & 3 & 4 & 0.9 \\
%\hline
%\end{tabular}
%\caption{Scalar parameters of depth \(d_t\), smoothness \(s_t\), and compression \(c_t\) describing each type of terrain \(t\).}
%\label{tab:terrain_values}
%\end{table}
%
%\begin{table}[h]
%\centering
%\begin{tabular}{ |c|c|c|c|c| } 
%\hline
%Terrain & Dry sand & Soil & Mud & Snow \\
%\hline
%$d_t$ & 0.5 & 0.4 & 2 & 3 \\ 
%$s_t$ & 3 & 2 & 2 & 4  \\ 
%$c_t$ & 0.2 & 0.6 & 0.6 & 0.9 \\ 
%\hline
%\end{tabular}
%\caption{Scalar parameters of depth \(d_t\), smoothness \(s_t\), and compression \(c_t\) describing each type of terrain \(t\).}
%\label{tab:terrain_values}
%\end{table}
%

The grass was represented as billboards (see Figure~\ref{fig:vegetation} and the accompanying video), and its deformation was implemented in Unity using shader-graphs to allow efficient space deformations. 
%Fig.~\ref{fig:grassComparison} shows the effect of the grass on the walking gait. Note in particular the right leg bending related to the effect of the virtual platform. 
We considered two typical grass heights : small ($0.5$ meters) and medium ($0.9$ meters). The size of the vegetation influences the maximal height of the platform set at one third of the vegetation's height.
We also forced the character to swing less in the case of higher vegetation, to model friction.
To this end, we adapted the controller parameter \(\beta\) to range linearly from 6 (no vegetation) to 4 (medium vegetation).

More examples combining different terrains, slopes and vegetation are shown in Fig.~\ref{fig:teaser} 
and in the supplemental video. In all situations shown, the character automatically adapts its gait while dynamically interacting with the surrounding, loose environment. 

\section{Limitations and Future Work}
%Discussion, Future Work and Conclusion}
%% MP: Having explicitly the word "limitation" is really important: Else the reviewers may say we overclaim, ant it will help Eduardo adding new contributions to complement this work.

This work shows that 
%illustrates how the use of 
simple layers acting at different levels 
%of detail of an animated character 
of motion controllers can enrich real-time character animation, and handle two-ways interactions with natural environments.
Simple and efficient, our method 
%is fully compatible with real-time animation,
achieves consistent changes of the environment and the character's gait.

Among limitations, sand footprints do not account for slope, leading to symmetrical shapes even on steep dunes where sand-slides should be taking place. Secondly, our handling of vegetation is limited, billboards being unable to capture realistic deformations. We also fail to handle different natures and densities of vegetation, such as plastic deformations for broken rods.
Lastly, our prototype implementation only models straight walks on small terrains. Extending balance control to orthogonal motion directions would allow the character to turn, while using adaptive, local subdivision of the terrain would allow to freely navigate the character on larger and visually richer terrains.

An interesting avenue for future work would be the addition of new control layers to other parts of the character to improve both its global and local adaptations to the environment. We are particularly interested in extending the approach to other part of the character such as the motion of his arms through high vegetation to clear bushes or protecting its head. Extension toward more accurate impact on the environment could be explored through learning local deformation that could be obtained from simulation or real data, while reinforcement learning could be used to improve the response on some layer of the character such as its feet placement.
%(as we observed on videos), 
Lastly, our method was specifically designed for walking motions on quite stable terrains.Generalizing control and interaction layers could enable to handle more complex cases, such as a character or any other creature, running, jumping or climbing a scree with rolling stones.

%\cp{Although we have demonstrated the simplicity and efficiency of our method, there's a lot of space for future work and improvements. First, our technique does not include the detection of the ground material, which makes it impossible for a character to adapt its motion dynamically to a terrain composed of different materials. The grass density is not detected either, which could help make the simulation more realistic. Second, we have limited our work to biped characters. This technique could be extended to four legged creatures, with an adaptation of the controller. Finally, the coefficients of the controller could be fine tuned, by using some motion capture data for example, and the elastic behavior of short grass could be improved.}
\section{Acknowledgments}
This work has been funded by the Marie Sklodowska-Curie Innovative Training Network CLIPE, in the framework of Horizon 2020.

% bibtex
\bibliographystyle{eg-alpha-doi} 
\bibliography{biblio}       

\newcommand{\etalchar}[1]{$^{#1}$}
\begin{thebibliography}{\uppercase{PALvdP18}}

\bibitem[BTZZ18]{Bermudez2018}
\textsc{Bermudez L., Tessendorf J., Zimmermann D., Zordan V.}:
\newblock {Real-time locomotion with character-fluid interactions}.
\newblock \emph{MIG 18} (2018).

\bibitem[HKS17]{HKS17}
\textsc{Holden D., Komura T., Saito J.}:
\newblock Phase-functioned neural networks for character control.
\newblock \emph{{ACM} Trans. Graph. 36}, 4 (2017).

\bibitem[KH10]{Kwon2010}
\textsc{Kwon T., Hodgins J.}:
\newblock {Control Systems for Human Running using an Inverted Pendulum Model
  and a Reference Motion Capture Sequence}.
\newblock \emph{SCA} (2010).

\bibitem[KLP20]{Kwon2020}
\textsc{Kwon T., Lee Y., Panne M. V.~D.}:
\newblock {Fast and flexible multilegged locomotion using learned centroidal
  dynamics}.
\newblock \emph{ACM Trans. Graph.} (2020).

\bibitem[MAA{\etalchar{*}}09]{MAA09}
\textsc{Mitake H., Asano K., Aoki T., Salvati M., Sato M., Hasegawa S.}:
\newblock Physics-driven multi dimensional keyframe animation for
  artist-directable interactive character.
\newblock \emph{Comput. Graph. Forum 28}, 2 (2009).

\bibitem[PALvdP18]{PALvdP18}
\textsc{Peng X.~B., Abbeel P., Levine S., van~de Panne M.}:
\newblock Deepmimic: Example-guided deep reinforcement learning of
  physics-based character skills.
\newblock \emph{ACM Trans. Graph. 37} (2018).

\bibitem[SM01]{SM01}
\textsc{Sun H.~C., Metaxas D.~N.}:
\newblock Automating gait generation.
\newblock In \emph{ACM SIGGRAPH} (2001).

\bibitem[SOH99]{Sumner99}
\textsc{Sumner R.~W., O'Brien J.~F., Hodgins J.~K.}:
\newblock {Animating Sand, Mud, and Snow}.
\newblock \emph{Comput. Graph. Forum} (1999).

\bibitem[WGH20]{Won2020}
\textsc{Won J., Gopinath D., Hodgins J.}:
\newblock {A Scalable Approach to Control Diverse Behaviors for Physically
  Simulated Characters}.
\newblock \emph{ACM Trans. Graph.} (2020).

\bibitem[YLvdP07]{SIMBICON}
\textsc{Yin K., Loken K., van~de Panne M.}:
\newblock {SIMBICON:} simple biped locomotion control.
\newblock \emph{{ACM} Trans. Graph. 26}, 3 (2007).

\end{thebibliography}

% biblatex with biber
% \printbibliography               

\end{document}